\documentclass[titlepage,12pt]{utarticle}
\usepackage{amsmath,amssymb,amsthm,amscd,cite,epsfig}
\usepackage{latexsym,amsmath,amssymb,graphicx}
\usepackage[all]{xy}
\usepackage{epsf}
\xyoption{v2}
\xyoption{2cell}
\xyoption{dvips}
\CompileMatrices
\LaTeXdiagrams
\UseAllTwocells
\newdimen\tableauside\tableauside=1.0ex
\newdimen\tableaurule\tableaurule=0.4pt
\newdimen\tableaustep
\def\phantomhrule#1{\hbox{\vbox to0pt{\hrule height\tableaurule width#1\vss}}}
\def\phantomvrule#1{\vbox{\hbox to0pt{\vrule width\tableaurule height#1\hss}}}
\def\sqr{\vbox{%
  \phantomhrule\tableaustep
  \hbox{\phantomvrule\tableaustep\kern\tableaustep\phantomvrule\tableaustep}%
  \hbox{\vbox{\phantomhrule\tableauside}\kern-\tableaurule}}}
\def\squares#1{\hbox{\count0=#1\noindent\loop\sqr
  \advance\count0 by-1 \ifnum\count0>0\repeat}}
\def\tableau#1{\vcenter{\offinterlineskip
  \tableaustep=\tableauside\advance\tableaustep by-\tableaurule
  \kern\normallineskip\hbox
    {\kern\normallineskip\vbox
      {\gettableau#1 0 }%
     \kern\normallineskip\kern\tableaurule}%
  \kern\normallineskip\kern\tableaurule}}
\def\gettableau#1 {\ifnum#1=0\let\next=\null\else
  \squares{#1}\let\next=\gettableau\fi\next}
\numberwithin{equation}{section}

\newcommand{\be}{\begin{equation}}
\newcommand{\ee}{\end{equation}}

\newcommand{\IP}{\mathbb{P}}

\newcommand\IZ{\mathbb {Z}}

\newcommand{\IR}{\mathbb{R}}
\newcommand{\ba}{\begin{array}}
\newcommand{\ea}{\end{array}}

\newcommand{\IF}{{\mathbb F}}

\newcommand{\kah}{{K\"ahler}}
\newcommand{\CE}{{\cal E}}

\newcommand{\bal}{\begin{aligned}}
\newcommand{\eal}{\end{aligned}}

\newcommand{\wq}{{\widetilde {Q}}}
\newcommand{\tp}{{\widetilde {P}}}
\newcommand{\CZ}{{\mathcal Z}}
\newcommand{\hc}{{\mathrm{ch}}}
\newcommand{\half}{{1\over 2}}

\begin{document}
\preprint{
    {\tt hep-th/0610068}\\
    SU-ITP-06/27\\
    SLAC-PUB-12147\\
}
\title{Black String Entropy and Fourier-Mukai Transform}
\author{Iosif Bena$^{\flat}$\footnote{{\tt iosif.bena@cea.fr}}~,
Duiliu-Emanuel Diaconescu$^{\$ }$\footnote{{\tt duiliu@physics.rutgers.edu}}~
and Bogdan Florea$^{\sharp}$\footnote{{\tt bflorea@slac.stanford.edu}}}
\oneaddress{
     {\centerline  {$^{\flat}$\it Service de Physique Th\'eorique, CEA/Saclay,}}
\smallskip
\centerline{\it F-91191 Gif-sur-Yvette Cedex, FRANCE}
    \smallskip
     {\centerline  {$^{\$ }$\it Department of Physics and Astronomy, 
     Rutgers University,}}
     \smallskip
     {\centerline {\it Piscataway, NJ 08854-0849, USA}}
      \smallskip
      {\centerline {$^{\sharp}$ \it Department of Physics and SLAC, 
Stanford University,}}
      \smallskip
      {\centerline {\it Palo Alto, CA 94305, USA}}
}
\date{October 2006}

\Abstract{
We propose a microscopic description of black strings 
in F-theory based on string duality and Fourier-Mukai transform. 
These strings admit several different microscopic descriptions 
involving D-brane as well as M2 or M5-brane configurations on elliptically 
fibered Calabi-Yau threefolds. In particular our results can also 
be interpreted as an asymptotic microstate count for D6-D2-D0 
configurations in the limit of large D2-charge on the elliptic fiber.
The leading behavior of the microstate degeneracy in this  
limit is shown to agree with the macroscopic 
entropy formula derived from the black string supergravity solution. 
} 

\maketitle

\section{Introduction} 

Black hole microstate counting has been a problem of constant  
interest in string theory \cite{Strominger:1996sh,
Dijkgraaf:1996cv, Breckenridge:1996is,Breckenridge:1996sn,Maldacena:1997de,
Vafa:1997gr,Maldacena:1999bp,Katz:1999xq,Bertolini:2001ns} 
for the past decade.  
This problem has been the subject of intense recent activity 
\cite{Vafa:2004qa,Aganagic:2004js,Dabholkar:2004yr,Dabholkar:2005qs,
Pioline:2005vi,Dabholkar:2005dt,Dabholkar:2005by,
Shih:2005qf,Shih:2005uc,Gaiotto:2005rp,Bena:2005va,
Berglund:2005vb,Aganagic:2005wn,Kraus:2005vz,
Gaiotto:2006ns,Gaiotto:2006wm,Beasley:2006us,deBoer:2006vg}
motivated by the connection with topological strings proposed in 
\cite{Ooguri:2004zv} and by the correspondence between 4D black holes 
and 5D black holes \cite{Gaiotto:2005gf} and black rings \cite{Bena:2005ay,
Gaiotto:2005xt,Elvang:2005sa,Bena:2005ni}.

In $N=2$ string theory compactifications, supersymmetric black holes 
can be described in terms of D-branes wrapping supersymmetric cycles 
in the internal manifold. The black hole 
entropy is determined by the degeneracy 
of D-brane bound states with fixed topological charges. In the semiclassical 
approximation, D-brane bound states are associated to cohomology classes 
on the moduli space of classical supersymmetric configurations.
The macroscopic entropy formula is typically captured by the asymptotic 
growth of BPS degeneracies in the limit of large charges. 
This has been shown in \cite{Maldacena:1997de,Vafa:1997gr,
Gaiotto:2005rp} for D4-D2-D0 configurations on Calabi-Yau threefolds. 
Analogous results for D-brane configurations with nonzero D6-brane 
charge seem to be more elusive. 

In this paper we address the problem of counting the microstate degeneracy 
for D-brane configurations with nonzero D6-brane charge on elliptically 
fibered Calabi-Yau threefolds. A string duality chain described in section 
two shows that this system admits several different descriptions in terms 
of wrapped branes in F-theory, M-theory or IIA compactifications. 
In particular this duality chain predicts an equivalence of the D6-D2-D0 
system with a D4-D2-D0 configuration on the same Calabi-Yau
threefold, which can be recognized as a Fourier-Mukai transform along 
the elliptic fibers. This is discussed in detail in section three. 
Another incarnation of the D6-D2-D0 configuration
which will play an important role in this paper is a noncritical 
six-dimensional string obtained by wrapping D3-branes on  holomorphic 
curves in F-theory compactifications.

D-brane systems with D4-D2-D0 charges have a known microscopic 
CFT description \cite{Maldacena:1997de,Vafa:1997gr,Gaiotto:2005rp} 
which allows one to compute the 
asymptotic degeneracy of states in the limit of large D0 charge. 
Our goal is to compare the resulting entropy formula with 
a macroscopic computation performed in a low energy supergravity 
description. We will show in section four that a reliable 
macroscopic description in the limit of large D0 charge 
must be formulated in terms of black-string solutions 
of $N=1$ six dimensional supergravity. The resulting macroscopic 
entropy formula reproduces the macroscopic result including 
certain subleading corrections. 

The problem of microstate degeneracies for D6-D4-D2-D0 black holes 
is also addressed in the upcoming work \cite{DM} using split attractor 
flows. Although this seems to be a different approach than the Fourier-Mukai 
transform employed here, it would be interesting to understand 
the relation between these two methods.

{\it Acknowledgments.} We would like to thank Sujay Ashok, Eleonora
dell'Aquila, Frederik Denef, Paul Horja, Robert Karp and 
especially Greg Moore for stimulating 
discussions. D.E.D. has been partially supported by NSF grant 
PHY-0555374-2006. 
\\
\\
{\it Note added.} 
This paper has some partial overlap with 
\cite{Aspinwall:2006yk} which appeared when this work was 
close to completion.

\section{A Duality Chain for Elliptic Fibrations} 

In this section we construct a duality chain involving
string compactifications on elliptic fibrations and explain 
its implications for black hole physics. 

The starting point of our discussion is a six dimensional
compactification of F-theory on an elliptically fibered Calabi-Yau threefold 
$X$ with a section. 
Consider a six-dimensional noncritical string obtained by wrapping a 
D3-brane on a smooth curve $C$ in $B$. Such strings have been studied 
extensively in the string duality literature especially  
when $C$ is a $(-1)$ curve on $B$
\cite{Klemm:1996hh,Mayr:1996sh,Lerche:1996ni,Minahan:1997ch,Minahan:1997ct,
Minahan:1998vr}. In this case they are related to tensionless strings 
associated to small instanton transitions in heterotic M-theory. 
Here we will take $C$ to be an arbitrary smooth curve in the base 
$B$. We will shortly see that $C$ must actually be a very ample divisor 
on $B$ of high degree. 

Six-dimensional noncritical strings are related by string duality to   
five-dimensional BPS particles or strings \cite{Vafa:1997gr}. 
The five-dimensional compactification of the theory on a circle 
of radius $R$ is equivalent to an M-theory compactification on 
$X$, where the size of the elliptic fiber is proportional to 
$1/R$. The six dimensional string can yield two types of objects, 
depending on its position relative to the compactification circle. 
A string wrapped on the compactification circle is equivalent to an 
M2-brane wrapping the horizontal curve $C$ in $X$. Moreover string
excitations carrying Kaluza-Klein momentum $n$ along the circle 
are dual to M2 bound states wrapping curves in the class $C + nF$, 
where $F$ is the class of the elliptic fiber. For curves $C$ of large 
degree and large $n$, such particles are expected to 
form five dimensional spinning 
black holes \cite{Vafa:1997gr,Katz:1999xq}. 
 
On the other hand an unwrapped string corresponds to an M5-brane 
wrapping the vertical divisor $D = \pi^{-1}(C)$ in $X$ (that is the 
complex surface obtained by restricting the elliptic fibration to $C$.)  
This is a five dimensional noncritical string. Let us further compactify the 
theory on an extra circle and consider wrapped noncritical strings
giving rise to four-dimensional particles.  
The excitations of the noncritical string 
with $n$ units of Kaluza-Klein momentum along the extra circle
correspond to D-brane bound 
states in the resulting IIA compactification on $X$. We obtain $n$ D0-branes 
bound to a D4-brane wrapping $D$. This picture has been employed in the 
microstate degeneracy counting of 
\cite{Vafa:1997gr}.

Motivated by the 4D/5D black hole correspondence, let us now 
consider an F-theory background of 
the form $X \times TN_{r} \times S^1 \times \IR$ 
where $TN_r$ is a Taub-NUT space of type $A_{r-1}$, 
and $\IR$ is the time direction.  This theory contains six-dimensional 
noncritical strings obtained as above by wrapping a D3-brane on 
$C\times S^1$, where $C$ is a curve in $B$. Note that the $SO(4)$ 
rotational symmetry of the transverse space to the string
is broken to $SU(2)$ by the Taub-NUT background. In the following we 
will be interested in spinning strings which carry angular 
momentum $J$ with respect to the unbroken $SU(2)$. 
Let us describe two sequences of duality transformations 
starting with this model. 

{\bf Sequence A.} 
Let us make the radius $R$ of the circle very small. 
Then we obtain an M-theory background $X\times TN_r \times \IR$ and 
the wrapped D3-brane is dual to an M2-brane wrapping $C$. The excitations 
of the string with $n$ units of KK momentum on $S^1$ correspond again to bound 
states of M2-branes wrapping a curve in the class $C+nF$. The duality
transformation preserves the spin quantum number $J$. Therefore a 
spinning string with angular momentum $J$ will be mapped to a 
spinning membrane configuration with the same spin quantum number. 
The resulting M-theory configuration falls in the class of models 
proposed in \cite{Gaiotto:2005gf}
and studied in detail in \cite{Behrndt:2005he}. 
We can further compactify this model along the $S^1$ fiber of the Taub-NUT 
space, obtaining a IIA compactification on $X$. The M2-brane configuration 
transverse to the Taub-NUT space corresponds to a D6-D2-D0 configuration on 
the same elliptic fibration $X$. The D6-brane has 
multiplicity $r$ and we have $m=2J$ units of 
D0-brane charge. This is the four dimensional limit of the 
correspondence considered in \cite{Gaiotto:2005gf,Behrndt:2005he}. 
However in our case the black hole interpretation of the above brane 
configurations involves some subtle aspects which will be discussed in section
four.

{\bf Sequence B.} Let us now perform a different chain of duality
transformations starting 
with the same F-theory configuration. Regarding this model as a IIB 
compactification on $B$, we will perform a T-duality transformation along 
the $S^1$ fiber of the Taub-NUT space followed by an M-theory lift. 
In spite of many subtleties, it is by now understood that T-duality 
along the fiber of the Taub-NUT space will give rise to  $r$ IIA 
NS five-branes wrapping $B$ and the remaining directions $S^1\times \IR$ 
\cite{Ooguri:1995wj,Diaconescu:1997gu,Gregory:1997te}. 
Note also that one of the transverse directions to the IIA NS5-branes 
is compactified on another circle ${\widetilde S}^1$, which is dual 
to the fiber of the Taub-NUT space.  
The D3-brane wrapping 
$C\times S^1$ is  mapped to a D4-brane wrapping
$C\times S^1\times {\widetilde S^1}$. A complete IIA description of the
model is quite awkward, and it is much more illuminating to take the M-theory 
lift instead. This yields an M-theory compactification on the elliptic 
threefold $X$.  The $r$ IIA NS5-branes lift to $r$ M5-branes 
wrapping the section $\sigma \simeq B$ of the elliptic fibration 
while the D4-brane 
lifts to an M5-brane wrapping the vertical divisor $D=\pi^{-1}(C)$. 
Both groups of M5-branes are wrapped on $S^1$. 
Note that the circle ${\widetilde S}^1$ together with the M-theory 
circle are now contained in the elliptic fiber of $X$. 
If $C$ is sufficiently ample the two groups of M5-branes can be deformed 
to a single smooth M5-brane wrapping a divisor in the linear 
system $r\sigma + \pi^*(C)$. The effective description of the 
M5-brane is a $(0,4)$ 
CFT on $\IR\times S^1$ as in \cite{Maldacena:1997de}.
One can further take the limit in which the radius of $S^1$ is very 
small obtaining a IIA D4-D2-D0 configuration on $X$. The D4-brane 
wraps a divisor in the class $(r\sigma + \pi^*C)$, and 
excitations of the original noncritical string with $n$ units of Kaluza-Klein 
momentum on $S^1$ are mapped to $n$ D0-branes bound to the D4-brane.  

The effect of this chain of duality transformations on the spin quantum 
number $J$ is harder to trace directly. However we will be able to compute it 
precisely once we identify the role of Fourier-Mukai transform in this
picture. 
Note that the combined effects of the above duality chains predict
a map between 
a D6-D2-D0 and D4-D2-D0 configurations on a given elliptic 
fibration $X$ resembling a T-duality 
transformation along the elliptic fiber. We will show in the next section
that such a map can be rigorously defined as a relative 
Fourier-Mukai transform. In particular, this will allow us to 
compute all charges of the D4-D2-D0 configuration including 
curvature corrections 
due to degenerate elliptic fibers.  

\section{Black Strings, D-Branes and Fourier-Mukai Transform} 

The goal of this section is to explain the relation between the 
Fourier-Mukai transform and the duality chain of section two, 
as well as its physical implications. 
As explained in the concluding remarks of the previous section, 
string duality predicts an equivalence of D6-D2-D0 and 
D4-D2-D0 configurations on the same elliptic Calabi-Yau 
threefold $X$. Since both sides of these equivalence are pure 
D-brane systems, we should be able to interpret this map as an 
autoequivalence of the derived category of $X$. 
In this section we will argue that the autoequivalence in question 
must be a relative Fourier-Mukai transform along the elliptic 
fibers. 
 
Our set-up is a IIA compactification on a smooth elliptically 
fibered Calabi-Yau threefold $X$. In order to set the ground 
for our discussion, we will start with a short review of
special \kah\ geometry, BPS states and D-branes.  
Throughout this paper we will identify the complexified 
\kah\ moduli space of $X$ with 
the complex structure moduli space of the mirror threefold 
$Y$. Let 
\be\label{eq:perA}
\Pi = \left[\CF_0,\CF_A, X^A, X^0\right]^{tr}
\ee
denote the periods of the holomorphic three-form on $Y$, 
where $A=1,\ldots,h^{1,1}(X)$. 
The inhomogeneous flat coordinates 
on the moduli space are 
\be\label{eq:perB}
t^A = {X^A\over X^0}, \qquad A=1,\ldots,h^{1,1}(X). 
\ee
The large radius limit point in the \kah\ 
moduli space of $X$ is identified with a large complex structure (LCS)
limit of $Y$. The periods are normalized 
so that $X^0$ is the fundamental period and $X^A$, 
are the logarithmic periods at the LCS point.  

The central charge of a BPS state with charges $(P^\Lambda,Q_\Lambda)$, 
$\Lambda=0,\ldots,h^{1,1}(X)$, 
is given by 
\be\label{eq:chchargeA} 
\CZ= e^{K/2}(Q_\Lambda X^\Lambda-P^\Lambda\CF_\Lambda) 
\ee
where 
\be\label{eq:kahpot}
K = - \mathrm{ln}\, i ({\overline X}^\Lambda \CF_\Lambda - X^{\Lambda} 
{\overline \CF}_\Lambda)
\ee
is the \kah\ potential. 

From a microscopic point of view, BPS states are bound states of 
D6-D4-D2-D0 brane configurations on $X$. Such configurations are 
described by holomorphic vector bundles, or, more generally, coherent on 
sheaves on $X$. Given such an object $\CE$, the central charge of the 
corresponding BPS state has an expansion of the form 
\be\label{eq:chchargeB} 
\CZ = e^{K/2} X^0 
\left(\int_X e^{J(t^A)}\mathrm{ch}(\CE) \sqrt{\mathrm{Td}(X)} +\ldots\right)
\ee 
near the large radius point, where $J(t^A)$ denotes the complexified 
\kah\ form on $X$, and $\ldots$ stand for 
world-sheet one-loop and instanton corrections. 
Following the standard conventions in the literature we will 
use the notation 
\be\label{eq:chchargeC} 
Z(\CE) = \int_X e^{J(t^A)}\mathrm{ch}(\CE) \sqrt{\mathrm{Td}(X)}.
\ee
More generally, if $\alpha$ is a cohomology class on $X$, we will 
denote by 
\be\label{eq:chchargeD} 
Z(\alpha) = \int_X e^{J(t^A)}\alpha \sqrt{\mathrm{Td}(X)}
\ee

Homological mirror symmetry implies that 
the logarithmic periods $X^A$ have an expansion of the 
form 
\be\label{eq:perC} 
X^A = X^0 (Z(\beta^A)+\ldots) 
\ee
near the large radius limit, where $\beta^A\in H^{2,2}(X)$ are 
Poincar\'e dual to some curve classes $C_A$ on $X$. 
The remaining periods have similar expansions 
\be\label{eq:perD}
\begin{aligned} 
\CF_A & = X^0 (Z(\CL_A) - Z(\CO_X)+\ldots)\cr 
\CF_0 & = X^0 (Z(\CO_X) +\ldots) \cr
\end{aligned}
\ee
for some holomorphic line bundles $\CL_A$ on $X$. 
We will denote by $\alpha_A = c_1(\CL_A) \in H^{1,1}(X)$. 
Moreover, $\{\alpha_A\}$ is a basis of $H^{1,1}(X)$, 
$\{\beta^A\}$ is a basis of $H^{2,2}(X)$, and 
\be\label{eq:perE} 
\int_X \alpha_A\wedge \beta^B = \delta_A^B.
\ee 

We can make a more specific choice of even cohomology generators 
taking into account the elliptic fibration structure of $X$. 
We will restrict ourselves to smooth elliptic fibrations 
$\pi:X\to B$ which can be written in Weierstrass form. 
The base $B$ is a smooth del Pezzo surface. 
Then $h^{1,1}(X) = h^{1,1}(B) +1$ and 
we can choose the basis $\{\alpha_A\}\subset H^{1,1}(X)$ so that 
\be\label{eq:basisA}
\alpha_i = \pi^*\gamma_i, \qquad i=1,\ldots h^{1,1}(B).
\ee
Moreover, the last basis element $\alpha_h$, where $h=h^{1,1}(X)$, 
is normalized so that 
\[
\int_F\alpha_h=1, \qquad \int_C \alpha_h = 0,
\] 
where $F$ denotes the class of the elliptic fiber, and $C$ is an 
arbitrary horizontal curve class\footnote{A curve class will be called 
horizontal if it lies in the image of the pushforward map 
$\iota_*:H_2(B) \to H_2(X)$, where $\iota:B \to X$ is the canonical 
section of the Weierstrass model.} on $X$. 
Denoting by $\sigma$ the 
$(1,1)$ class related by Poincar\'e duality to the section class, 
we have 
\be\label{eq:normcls}
\alpha_h = \sigma + \pi^* c_1(B).
\ee
Let $\{\eta^i\}$, $i=1,\ldots,h^{1,1}(B)$ denote the dual basis 
of $H^{1,1}(B)$, i.e. 
\be\label{eq:dualbasisA}
\int_B \gamma_i \wedge \eta^j = \delta_i^j.
\ee
Then we can choose the basis $\{\beta^A\}\subset H^{2,2}(X)$  
so that 
\be\label{eq:basisB}
\beta^i = \sigma \wedge \pi^*\eta^i,\qquad i=1,\ldots,h^{1,1}(B) 
\ee
and $\beta^h$ is Poincar\'e dual to the fiber class $F$. 

The D6-D2-D0 configurations related by duality to 
F-theory noncritical strings are described by 
holomorphic bundles $\CE$ on $X$ with 
Chern character
\be\label{eq:charges}
\mathrm{ch}(\CE) = r- \sum_{i=1}^{h^{1,1}(B)} q_i \beta^i -n\beta^h -
m \omega \in H^0(X) \oplus H^{2,2}(X) \oplus H^{3,3}(X) 
\ee 
where $\omega \in H^{3,3}(X)$ is the fundamental class of $X$
normalized so that 
\[
\int_X \omega =1.
\]
A straightforward computation shows that 
\be\label{eq:chchargeE} 
Z(\CE) = rZ(\CO_X) - q_i Z(\beta^i) - nZ(\beta^h) -m
\ee
This expression determines the charge vector of the corresponding 
BPS state 
\be\label{eq:chvectA}
(P^0,P^A,Q_A,Q_0) = (r,0,-q_i,-n,-m). 
\ee
As explained in section two, $r$ is the charge 
of the transverse Taub-NUT space in the F-theory model,
$C=q_i\eta^i$ is a horizontal 
curve on $X$ which is identified with the support $C\subset B$ 
of the wrapped D3-brane, and $m=2J$ is twice the angular momentum. 

The microscopic entropy of such a D-brane system is determined by counting 
cohomology classes on the moduli space of classical supersymmetric 
configurations. From a mathematical point of view, supersymmetric D-brane 
configurations correspond to semi-stable 
coherent sheaves on $X$ with fixed Chern classes given by \eqref{eq:charges}. 
In general the geometry of moduli spaces of semi-stable coherent sheaves 
is very little understood on Calabi-Yau threefolds. 
These spaces are expected to have very complicated singularities which 
make a mathematical formulation of the counting problem very difficult.

The D6-D2-D0 configurations considered in this section can however be 
mapped to D4-D2-D0 configurations by the duality chain of section two. 
We will show below that this map is in fact a relative 
Fourier-Mukai transform along the 
elliptic fibers. Then the counting problem becomes more tractable, and 
we can employ the methods of 
\cite{Maldacena:1997de,Vafa:1997gr,Gaiotto:2005rp} 
in order to determine the asymptotic growth of the microstates in the 
limit of large D2-brane charge on the elliptic fiber. 

The physical applications of the Fourier-Mukai transform have been focused
so far on heterotic bundle constructions and heterotic-F-theory duality 
starting with the work of 
\cite{Friedman:1997yq,Friedman:1997ih,Bershadsky:1997zs,Donagi:1998vw,
Donagi:1998vx}. 
It has also been considered in 
\cite{Andreas:2000sj,Andreas:2001ve} 
in connection with homological mirror symmetry, which is closer to 
our context. 
The Fourier-Mukai transform can be intuitively thought of as T-duality 
along the elliptic fibers. However naive T-duality is not well defined
in the presence of singular elliptic fibers, hence we have to employ a
more sophisticated transformation which is defined abstractly 
as a derived functor. Since the technical details have been thoroughly 
worked out in the above papers, we will only recall the essential facts 
omitting most technical details. It is worth noting however that 
the Fourier-Mukai transform is not an element of the T-duality
group of the theory, which is generated by monodromy transformations
acting on the derived category \cite{Horja:1999,Horja:2001}. 
This question was investigated 
in detail in \cite{Andreas:2000sj,Andreas:2001ve}, where it was found 
that the Fourier-Mukai transform differs from a monodromy transformation 
by a certain twist. 
This agrees with the transformation found in section two, which 
involved nonperturbative duality transformations.  

The Fourier-Mukai transform of the D6-D2-D0 configuration described 
by a bundle $\CE$ is a D4-D2-D0 system described by a
derived object $\CF[1]$, where $\CF$ is a torsion sheaf 
$\CF$ on $X$ supported on a divisor $\Sigma \subset X$.
The effect of the shift by $1$ is to change the sign of all 
D-brane charges of the configuration represented by the sheaf 
$\CF$.  
Moreover, according to \cite{Andreas:2003zb}, the Fourier-Mukai
transform preserves semi-stability with respect to a suitable 
polarization of $X$. This means it maps supersymmetric D-brane 
configurations to supersymmetric D-brane configurations, therefore 
we can reliably use it in order to count BPS states. 
 
The Chern character of $\CF$ is  \cite{Andreas:2000sj} 
\be\label{eq:FMtransA} 
\begin{aligned} 
\hbox{ch}_1(\CF) & = r \sigma + \pi^* C \cr
\hbox{ch}_2(\CF) & = -{r\over 2} \sigma \wedge \pi^*c_1(B) + 
\left(m+{1\over2} \int_X \sigma \wedge \pi^*c_1(B)
\wedge \pi^*C \right) \beta\cr
\hbox{ch}_3(\CF) & = -n\omega + {r\over 6} \sigma\wedge \pi^*c_1(B)^2.\cr
\end{aligned} 
\ee
where
\be\label{eq:curvecls}
C = q_i \eta^i. 
\ee
Note that $C$ can be interpreted by Poincar\'e duality as a curve 
class on $B$. We will assume that $C$ is 
a very ample divisor class on $B$ of sufficiently high degree so that 
the generic surface $\Sigma$ in the class $r\sigma + \pi^*C$ is smooth and 
irreducible. 

The action of the Fourier-Mukai transform on topological charges 
is in agreement with the map found in the previous section up 
to curvature corrections involving $c_1(B)$. This is positive 
evidence for the identification of these two transformations. 
The curvature corrections are not under control in the chain 
of dualities described in section two, hence we will not be able 
to perform a more detailed check. We will obtain more compelling 
evidence by matching the black hole entropy formulas in the next section. 

The leading 
contribution to the entropy of a D4-D2-D0 configuration 
in the limit of large D0 charge has been evaluated in  
\cite{Maldacena:1997de,Vafa:1997gr,Gaiotto:2005rp}. 
As a first step, we need to identify the BPS charges 
$({\widetilde P}^A, {\widetilde Q}_A)$ of this configuration
by computing the leading terms of the central charge 
\be\label{eq:shiftbyone}
Z(\CF[1])=-Z(\CF)
\ee 
near the large radius limit point. 
More precisely, we have to express
\be\label{eq:chvectB}
\begin{aligned} 
Z(\CF) & = \int_X e^J\mathrm{ch}(\CF) \sqrt{Td(X)}\cr
& = \int_X \left[{1\over 2}J^2 \hc_1(\CF) + J\hc_2(\CF) + {1\over 2}\hc_1(\CF) 
\mathrm{Td}_2(X) + \hc_3(\CF)\right]\cr
\end{aligned}
\ee
as a linear combination of the functions $Z(\CL_A)-Z(\CO_X)$
and $Z(\beta^A)$ which appear in the expansion \eqref{eq:perD} 
of the periods at the large radius limit point. 

For this computation we will need the triple intersection numbers 
\be\label{eq:tripleA}
\begin{aligned}
& D_{hhh} = {1\over 6} \int_X \alpha_h^3 = 
{1\over 6} \int_X(\sigma + \pi^*c_1(B))^3 = 
{1\over 6}\int_B c_1(B)^2
\cr
& D_{hh i} = {1\over 6} \int_X \alpha_h^2 \wedge \alpha_i = 
{1\over 6}\int_X \sigma \wedge \sigma \wedge \alpha_i 
= {1\over 6} \int_B c_1(B) \wedge \gamma_i \cr
& D_{h i j} = {1\over 6}\int_X \alpha_h \wedge \alpha_i \wedge \alpha_j 
={1\over 6}\int_X \sigma \wedge \alpha_i \wedge \alpha_j 
= {1\over 6} \int_B \gamma_i \wedge \gamma_j \cr
& D_{i j k } ={1\over 6}\int_X \alpha_i \wedge \alpha_j \wedge \alpha_k =0.\cr
\end{aligned}
\ee
Let us introduce the following notation 
\be\label{eq:notation}
d = \int_B c_1(B)^2 \qquad c_i = \int_B c_1(B) \wedge \gamma_i \qquad 
d_{ij} = \int_B \gamma_i \wedge \gamma_j.
\ee
Note that we have 
\be\label{eq:dualint} 
\int_B \eta^i \wedge \eta^j = d^{ij}\qquad \int_B c_1(B) \wedge \eta^i = 
d^{ij}c_j \qquad \hbox{where}\qquad  
d_{ik}d^{kl} = \delta_i^l. 
\ee
We will also make frequent use of the following expressions 
\be\label{eq:morestuff}
Q=\int_B C \wedge C = q_id^{ij}q_j \qquad 
c=\int_B C\wedge c_1(B) = q_id^{ij}c_j. 
\ee
The Chern character $\hc(\CF)$ written in terms of 
the bases $\{\alpha_A\}$ and $\{\beta^A\}$ reads  
\be\label{eq:chernB}
\begin{aligned} 
\hc(\CF)& = r\alpha_h + (q_i - rc_i) d^{ij}\alpha_j 
+\left(m+{c\over 2}\right) \beta^h-{r\over 2} c_i\beta^i -
\left(n-{rd\over 6}\right)\omega .\cr
\end{aligned}
\ee
Now we substitute equation \eqref{eq:chernB} in \eqref{eq:chvectB}
obtaining 
\be\label{eq:chvectC} 
\begin{aligned}
Z(\CF) & = 
\int_X \bigg[ {1\over 2}J^2 (r \alpha_h +(q_i-rc_i)d^{ij}\alpha_j)]
+ \left(m+{c\over 2}\right) J\beta^h - {r\over 2}c_iJ\beta^i \cr
& \qquad \quad 
-(n-{rd\over 6})\omega + {1\over 2} (r\alpha_h +(q_i-rc_i)d^{ij}\alpha_j)
\mathrm{Td}_2(X) \bigg]\cr
\end{aligned}
\ee
We have to express \eqref{eq:chvectC} as a linear combination of the 
functions 
\be\label{eq:chvectD}
\begin{aligned} 
Z(\CL_i)-Z(\CO_X) & = \int_X (e^{J+\alpha_i}-e^{J})\left(1+{1\over 2}
\mathrm{Td}_2(X)\right)
= \int_X \left[\half J^2 \alpha_i + \half J \alpha_i^2 + \half 
\alpha_i \mathrm{Td}_2(X) \right]\cr
Z(\CL_h)-Z(\CO_X) & = \int_X (e^{J+\alpha_h}-e^{J})\left(1+{1\over 2}
\mathrm{Td}_2(X)\right)
= \int_X \left[\half J^2 \alpha_h + \half J \alpha_h^2 + \half 
\alpha_h \mathrm{Td}_2(X) +{\alpha_h^3\over 6}\right]\cr
Z(\beta^i) & = \int_X e^J\beta^i \left(1+{1\over 2}
\mathrm{Td}_2(X)\right) = \int_X J\beta^i.\cr
Z(\beta^h) & = \int_X e^J\beta^h \left(1+{1\over 2}
\mathrm{Td}_2(X)\right) = \int_X J\beta^h.\cr
\end{aligned}
\ee
Taking into account equations \eqref{eq:notation}, 
\eqref{eq:dualint} and the following identity 
\[ 
\begin{aligned} 
\alpha_h^2 = (\sigma + \pi^*c_1(B))^2 = c_i \beta^i + d \beta^h 
\end{aligned}
\] 
we obtain
\be\label{eq:chvectE} 
\begin{aligned} 
Z(\CF) & = r(Z(\CL_h) -Z(\CO_X)) + (q_i-rc_i)d^{ij} (Z(\CL_i)-Z(\CO_X))\cr
&\quad + \left[m+\half (q_i-rc_i)d^{ij}(c_j-d_{jj})\right]Z(\beta^h)
-rc_iZ(\beta^i) -n\cr
\end{aligned}
\ee
Note that 
\[ 
c_j -d_{jj} = \int_B (c_1(B)-\gamma_j)\wedge \gamma_j = 
2(1- \chi(\CO_B(\gamma_i)))
\]
where $\CO_B(\gamma_i)$ denotes the holomorphic line bundle 
on $B$ determined by the divisor class $\gamma_i$. 
Then our final formula reads 
\be\label{eq:chvectF} 
\begin{aligned} 
Z(\CF) & = r(Z(\CL_h) -Z(\CO_X)) + (q_i-rc_i)d^{ij} (Z(\CL_i)-Z(\CO_X))\cr
&\quad + \left[m+(q_i-rc_i)d^{ij}(1-\chi(\CO_B(\gamma_j)))\right]Z(\beta^h)
-rc_iZ(\beta^i) -n.\cr
\end{aligned}
\ee
Taking into account the sign in equation \eqref{eq:shiftbyone}, 
we can now read off the charge vector of the corresponding 
BPS state 
\be\label{eq:chvectG} 
\begin{aligned}
& \left({\widetilde P}^0, {\widetilde P}^A, {\widetilde Q}_A, 
{\widetilde Q}_0\right)
= \left(0,(q_i-rc_i)d^{ij},r,-rc_i,
m+(q_i-rc_i)d^{ij}(1-\chi(\CO_B(\gamma_j))),n\right).\cr
\end{aligned}
\ee

According to \cite{Maldacena:1997de}, the asymptotic 
microstate degeneracy of the D4-D2-D0 system in the limit of 
large D0 charge is determined by the degeneracy of states in a 
$(0,4)$ CFT obtained by lifting the system to M theory. 
The left moving central charge of the CFT is given by 
\be\label{eq:leftcharge} 
c_L = D+{1\over 6}\int_X (r\sigma + \pi^*C) \wedge c_2(X) 
\ee
where 
\be\label{eq:tripleB}
\begin{aligned}
D & = {1\over 6}\int_X (r\sigma + q_i \pi^* \eta^i)^3 = 
{1\over 6}\left(dr^3-3r^2q_ic_id^{ij}
+3rq_iq_jd^{ij}\right)\cr
& = {1\over 6}(dr^3 - 3r^2c + 3rQ).\cr
\end{aligned}
\ee
The microstate degeneracy is determined by the  
asymptotic growth of states of momentum 
\be\label{eq:effdzero}
{\widehat m } = n + {1\over 12}
(D^{\alpha\alpha} \wq_{\beta}\wq_{\beta} 
+ 2 D^{\alpha i} \wq_\beta \wq_i + D^{ij} \wq_i\wq_j)
\ee 
where 
\[
\left[
\begin{array}{cc} 
D^{\alpha \alpha} & D^{\alpha i} \cr
D^{i\alpha } & D^{ij} \cr
\end{array} 
\right] = 
\left[
\begin{array}{cc} 
D_{\alpha \alpha} & D_{\alpha i} \cr
D_{i\alpha } & D_{ij} \cr
\end{array} 
\right]^{-1}.
\] 
and 
\be\label{eq:contraction} 
\begin{aligned} 
D_{\alpha\alpha} & = D_{\alpha\alpha\alpha}\tp^{\alpha} + 
D_{\alpha\alpha i} \tp^i = {1\over 6}(rd+(q_i-rc_i)d^{ij}c_j) =
{1\over 6} q_ic_jd^{ij}\cr
D_{\alpha i} & = D_{\alpha i \alpha} \tp^\alpha + D_{\alpha i j} \tp^j 
= {1\over 6}(rc_i + d_{ij}(q_k-rc_k)d^{kj}) = {1\over 6}q_i\cr
D_{i j } & = D_{ij\alpha} \tp^\alpha = {1\over 6} rd_{ij} \cr
\end{aligned} 
\ee  
Applying Cardy's formula, we find the leading term in the entropy 
formula to be 
\be\label{eq:microent}
S_{micro}= 2\pi \sqrt{D{\widehat m}}.
\ee
Note that this formula captures the microstate degeneracy due to a 
gas of ${\widehat m}$ D0-brane bound to a fixed D4-brane wrapping a
divisor $\Sigma$ in the class $(r\sigma + \pi^*C)$
\cite{Gaiotto:2005rp}. 
In particular, this is not an exact formula for the entropy of the 
D4-D2-D0 configuration, and it does not capture the asymptotic 
behavior at large $r$. In order to capture the later behavior 
one has to integrate on the moduli space of the D4-brane, which is 
a very difficult computation. We leave this issue for later work. 

\section{Six Dimensional Black Strings and Macroscopic Entropy}  

The purpose of this section is to find the macroscopic description of the 
brane configurations discussed in sections two and three in terms 
of low energy supergravity. We will first show that the four or 
five dimensional attractor mechanism does not yield reliable 
solutions in the limit required by Cardy's formula. We will also 
show that a reliable low energy description of the system must be 
formulated in terms of black string solutions of six dimensional 
$N=1$ supergravity. The black string entropy will be shown to agree with 
the leading behavior of the microscopic result 
\eqref{eq:microent} in the limit of large charges.

\subsection{D6-D2-D0 Attractors} 

Let us first try to solve the attractor equations 
\cite{Ferrara:1996dd,Ferrara:1996um,Shmakova:1996nz} for black holes 
carrying D6-D2-D0 charges in a neighborhood of the large 
radius limit point in the \kah\ moduli space. According to 
\cite{Gaiotto:2005gf,Behrndt:2005he}, this is equivalent to 
solving five dimensional attractor equations for the 
dual M2-brane configurations.

Following \cite{Behrndt:1996jn,Shmakova:1996nz}, we write the 
attractor equations in the form 
\be\label{eq:attractA} 
\begin{aligned} 
iP^\Lambda & = Y^\Lambda - {\overline Y}^\Lambda \cr
iQ_\Lambda & = \CF_\Lambda(Y) - {\overline \CF}_\Lambda(Y) \cr
\end{aligned}
\ee
where the new variables $Y^\Lambda$ are defined by 
\[ Y^\Lambda = {\overline \CZ} X^\Lambda.\] 
Here $\CZ$ denotes the central charge of a BPS states with 
charges $(p^\Lambda,q_\Lambda)$ \eqref{eq:chchargeB}.
The macroscopic entropy is given by  
\be\label{eq:macroentA} 
S_{macro} = i\pi \left({\overline Y}^\Lambda \CF_\Lambda(Y) 
- Y^\Lambda {\overline \CF}_\Lambda(Y)\right).
\ee
In our case the charge vector is given by \eqref{eq:chvectA}, hence 
the equations \eqref{eq:attractA} reduce to 
\be\label{eq:attractB} 
\begin{aligned} 
iP^0 & = Y^0 -{\overline Y}^0\qquad 
&iQ_0 & = \CF_0(Y) - {\overline \CF}_0(Y) \cr
0 & = Y^A - {\overline Y}^A\qquad 
&iQ_A & = \CF_A(Y) - {\overline \CF}_A(Y) \cr
\end{aligned}
\ee
The solution of these equations is of the form 
\cite{Shmakova:1996nz,Moore:1998pn}
\be\label{eq:attractsolA}
\begin{aligned} 
S_{macro} & = {\pi\over 3P^0} 
\sqrt{{4\over 3}(\Delta_Ay^A)^2 -9{({(P^0)}^2Q_0)}^2}\cr
t^A & = {3\over 2} {y^A\over \Delta_Ay^A}(P^0Q_0) -
i{3\over 2} {y^A\over \Delta_A y^A} {S_{macro}\over \pi}\cr
\end{aligned}
\ee
where $y^A$ are solutions to the quadratic equations
\be\label{eq:attractsolB} 
D_{ABC} y^Ay^B = \Delta_C, \qquad \Delta_C=-P^0Q_C.
\ee
An existence condition for the attractor point is that the 
solutions $y^A$ of \eqref{eq:attractsolB} be real. 
Moreover the attractor solution is self-consistent only if 
the imaginary parts $\mathrm{Im}(t^A)$ of the \kah\ 
parameters in \eqref{eq:attractsolA} are large and negative. 

Next let us specialize equations \eqref{eq:attractB} to 
D6-D2-D0 configurations on elliptic Calabi-Yau threefolds. 
In this case, the 
charge vector is given in equation 
\eqref{eq:chvectA}. We find that the entropy formula is given by 
\be\label{eq:entropyA} 
S_{macro} = {\pi\over 3r} 
\sqrt{{4\over 3} (\Delta_h y^h + \Delta_i y^i)^2 -9r^4m^2}
\ee
where 
\be\label{eq:entropyB} 
\Delta_h =  rn\qquad 
\Delta_i = r q_i 
\ee 
and $y^i,y^h$ are solutions of the system of quadratic equations 
\be\label{eq:entropyC}
\begin{aligned} 
D_{hhh} (y^h)^2 + 
2 D_{hhi}y^h x^i + D_{ijh} y^iy^j & = rn \cr
D_{hhi}(y^h)^2 + 2 D_{i j h} y^h y^j 
& = r q_i. \cr
\end{aligned}
\ee
Using formulas \eqref{eq:tripleA}, \eqref{eq:notation}, 
equations \eqref{eq:entropyC} become 
\be\label{eq:entropyD} 
\begin{aligned} 
{1\over 6} d(y^h)^2 + {1\over 3} c_i y^h y^i + {1\over 6} 
d_{ij}y^iy^j & = rn\cr
{1\over 6} c_i (y^h)^2 + {1\over 3} d_{ij} y^h y^j &
= r q_i \cr
\end{aligned}
\ee
Using the linear equations in the $y^i$, we find 
\be\label{eq:entropyE}
y^i = d^{ij}\left({3rq_j\over y^h} - {c_j y^h\over 2}\right)
\ee
Substituting equations \eqref{eq:entropyE} in the first equation in 
\eqref{eq:entropyD} 
we obtain 
the quartic equation  
\be\label{eq:entropyG} 
{d\over 24} (y^h)^2 + {r\over 2} q_i d^{ij} c_j + {3r^2 \over 2} 
q_id^{ij}q_j (y^h)^{-2} = rn. 
\ee 
Using the notations \eqref{eq:morestuff}, we 
can rewrite equation \eqref{eq:entropyG} in the final form 
\be\label{eq:entropyH} 
{d\over 24} (y^h)^4 -r\left(n-{c\over 2}\right) 
(y^h)^2 + {3r^2Q\over 2} =0.
\ee
Solving for $(y^h)^2$, we find 
\be\label{eq:entropyI} 
(y^h_\pm)^2 = 
{12r\over d} \left[\left(n-{c\over 2}\right) \pm 
\sqrt{ \left(n-{c\over 2}\right)^2 - {dQ\over 4}}\right]
\ee
Using equations \eqref{eq:entropyE}, the macroscopic entropy 
formula \eqref{eq:entropyA} can be expressed as a function of $y^h$ 
as follows 
\be\label{eq:entropyJ} 
S_{macro} = 
{\pi \over 3} \sqrt{{4\over 3}\left[\left(n-{c\over 2}\right)
y^h_\pm +{3rQ\over y^h_\pm}\right]^2 -9r^2m^2}
\ee
The values of the \kah\ moduli at the attractor point are given by 
\be\label{eq:kahparamB}
\begin{aligned} 
t^h & = {3rm  y^h \over 2\Delta} - i 
{3 y^h S_{macro}\over 2\pi\Delta } \cr
t^i & = {3rm y^i\over 2\Delta} -i 
{3 y^i S_{macro}\over 2\pi\Delta}\cr
\end{aligned}
\ee
where 
\be\label{eq:kahmodB}
\begin{aligned} 
\Delta & = 
\left(n-{c\over 2}\right) y^h_\pm+ {3rQ\over y^h_\pm}\cr
\end{aligned}
\ee
Now let us review the regime of validity of the microscopic 
formula \eqref{eq:entropyJ}. We must satisfy the following conditions
\begin{itemize} 
\item[$(i)$] The curve class $C=q_i\eta^i$ should be sufficiently ample 
on $B$ so that $\Sigma = r\sigma + \pi^*C$ is a very ample divisor 
on $X$. More precisely, a generic surface $\Sigma$ is smooth 
and irreducible if $C$ is an effective curve class on $B$ 
and also $C-c_1(B)$ is a smooth irreducible curve on $B$
\cite{Tony}. If we choose the basis elements $\eta^i$, $i=1,\ldots,
h^{1,1}(B)$, to be Poincar\'e dual to generators of the Mori cone, 
the first condition implies that the integers $q_i$ must be positive. 
The second condition implies that $q_i>rc_i$ 
for all $i=1,\ldots,h^{1,1}(B)$. Note that Cardy's formula for 
the entropy becomes more and more reliable as we increase $q_i$ 
keeping $r$ fixed because the divisor $\Sigma$ becomes more and 
more ample. Then one can neglect the effect of singular divisors 
in the linear system $|\Sigma|$ on the target space geometry of the 
$(0,4)$. As shown in \cite{Minasian:1999qn}, the $(0,4)$ sigma model 
for the M5-brane is quite involved, and the effects of singular 
divisors are not under analytic control. We expect these effects to 
become important for values of $q_i$ comparable to $r$. In particular, 
if $q_i<rc_i$, the divisor $\Sigma$ is not smooth, and the $(0,4)$ 
description employed in \cite{Maldacena:1997de} breaks down. 

\item[$(ii)$] Assuming condition $(i)$ to be satisfied, validity of 
Cardy's formula also requires the momentum of the CFT states 
to be much larger than the central charge. 
This condition is satisfied if the D0-brane 
charge $n$ is much larger than $D=\Sigma^3$. From the point of view of the 
D4-D2-D0 configuration discussed in section three, this means that the 
formula \eqref{eq:microent} captures the asymptotic behavior of 
the microstate degeneracy in the limit of large $n$ keeping $r,q_i$ 
fixed. 
\end{itemize}

\noindent 
The two solutions found in 
\eqref{eq:entropyI} have the following leading order behavior in 
the limit of large $n$, with $r,q_i$ fixed
\be\label{eq:attractH}
(y^h_+)^2 \sim {24rn\over d} \qquad 
(y^h_-)^2 \sim {3Q\over 2n}. 
\ee
One can rule out the first solution observing that for any choice of 
the sign for $y_h$ at least one of the attractor \kah\ parameters 
is large and positive\footnote{We thank F. Denef and G. Moore for 
discussions on this point.}.
This is incompatible with a physical interpretation 
of the solution, since it would require a large negative volume 
of the Calabi-Yau threefold.
 
For the second solution, the leading term of the 
macroscopic entropy formula is 
\be\label{eq:largeentA}
S_{macro} \sim \pi \sqrt{2rnQ-r^2m^2} 
\ee
and the leading behavior of the \kah\ moduli at the attractor point 
is 
\be\label{eq:kahparamC} 
\begin{aligned} 
\mathrm{Im}(t^h) & \sim - \sqrt{rQ\over 2n}\cr
\mathrm{Im}(t^i) & \sim - d^{ij}\left[q_j\sqrt{2n\over rQ} 
-{c_j\over 2}\sqrt{rQ\over 2n}\right].
\end{aligned} 
\ee
Clearly, for large $n$, $\mathrm{Im}(t^i)$ are very large and negative
while $\mathrm{Im}(t^h)$ is negative but very small. Such points do not lie 
in the neighborhood of the large radius limit of the \kah\ moduli 
space, hence the attractor solution is not self-consistent. 
One may wonder if a self-consistent attractor solution may exist in 
other regions of the moduli space. The quantum special geometry of the 
\kah\ moduli space has been solved for the elliptic fibration 
over $\IP^2$ in \cite{candelas:1994hw}. Their results show that there 
is no region in the moduli space where the quantum area of the elliptic 
fiber is much smaller than the quantum area of a horizontal 
curve. This does not logically rule out the existence of 
attractor points in quantum phases of the moduli space,
but it suggests that this would not be a natural solution to our 
problem. 

In the following we would like to propose another resolution of 
this problem suggested by the duality chain of section two. 
Note that according to \cite{Aspinwall}, IIA compactifications 
on elliptic fibrations are equivalent to six dimensional F-theory 
compactifications on the base in the limit of small elliptic fibers.
In this limit, a D6-D2-D0 configuration is mapped to a D3-brane 
wrapping a holomorphic curve in the base, as discussed in section two. 
The resulting noncritical string also wraps a transverse circle $S^1$ 
whose radius is inversely proportional with the size if the elliptic 
fiber. 
Therefore the scaling behavior of the \kah\ parameters at the 
attractor point suggests that the correct low energy description 
of our system should be formulated in terms of black string 
solutions of $N=1$ six dimensional supergravity. 

\subsection{Black Strings in $N=1$ supergravity} 

Let us start with a brief review of F-theory compactifications 
to six dimensions from the low energy point of view. 
Since we are interested only in compactifications on smooth 
Weierstrass models $X\to B$, we have to take $B$ to be a 
smooth Fano surface, i.e. a del Pezzo surface. Therefore $B$ 
can be either a $k$-point blow-up of $\IP^2$, $0\leq k \leq 8$ 
or $\IF_0=\IP^1\times \IP^1$. For future reference we will 
choose a basis $\{\gamma_i\}$ of $H^{1,1}(B)$ of the form 
\be\label{eq:oneonebasisA} 
\begin{aligned} 
& \gamma_1 = e_1&  &\quad&  & \gamma_2 = e_2 & 
&\quad&  &\ldots&  
&\quad&  &\gamma_{k} = e_k& & \quad&  & \gamma_{k+1} =h &  &\quad & &\mathrm{for} \
B = dP_k&  \cr
& \gamma_1 = a&  &\quad&  & \gamma_2 = b & 
&\quad&  & &  
&\quad&  & & & \quad&  &  & &\quad & &\mathrm{for} \
B = \IF_0&  \cr
\end{aligned}
\ee
where $h$ denotes the hyperplane class of $\IP^2$ and $e_1,\ldots, e_k$ 
denote the exceptional curve classes. 
In the case $B=\IF_0$, $a,b$ denote the classes of the two rulings. 
The dual basis $\{\eta^i\}$ is 
\be\label{eq:oneonebasisB} 
\begin{aligned} 
& \eta^1 = -e_1&  &\qquad&  & \eta^2 = -e_2 & &\quad&  &\ldots&  
&\quad&  &\eta^{k} = -e_k&  &\quad&  
&\eta^{k+1} = h&  & \quad&  &\mathrm{for} \
B = dP_k&  \cr
& \eta^1 = b&  & \quad&  & \eta^2 =a&  & & & &  
& &  & &  & &  
& &  & & & \mathrm{for} \
B = \IF_0&  \cr
\end{aligned}
\ee
The intersection matrix $(d_{ij})$ reads
\be\label{eq:oneonebasisC}
\begin{aligned}
& (d_{ij})=\mathrm{diag}(-1,-1,\ldots,-1,1)& &\quad& &\mathrm{for}\ B=dP_k& \cr
& (d_{ij})=\left(\begin{array}{cc}0 & 1\cr 1 & 0 \cr
\end{array}\right)& &\quad& &\mathrm{for}\ B=\IF_0& \cr
\end{aligned}
\ee

Let us first consider the case $B=dP_k$, $0\leq k\leq 8$, 
leaving $B=\IF_0$ for a separate discussion.  
The low energy supergravity theory contains a $N=1$ graviton multiplet and 
$k=(h^{1,1}(B)-1)$ $N=1$ tensor multiplets. The bosonic spectrum consists of 
the metric tensor, $(k+1)$ elementary tensor multiplets 
and $k$ real scalar fields.   
The tree level formulation of the theory  
has been described in detail in \cite{Romans:1986er}. 
The scalar components of the tensor multiplets take values in the 
coset manifold $O(k,1)/O(k)$. They are parameterized by 
an $O(k,1)$ valued field 
\[ 
V(x) = \left[ \begin{array}{cc} x^{a}_{b} & x^a_{k+1} \\ v_a & v_{k+1} \\
\end{array} \right] 
\]
where $a,b=1,\ldots,k$
subject to local $O(k)$ gauge transformations 
\[ V(x) \to g(x) V(x),\qquad g(x)\in O(k)\] 
and global $SO(1,k)$ symmetry transformations 
\[V(x) \to V(x)U^{-1},\qquad  U\in O(k,1).\]
The $(k+1)$ elementary antisymmetric 
tensor fields $B^1,\ldots,B^{k+1}$ are obtained by Kaluza-Klein 
reduction of the type IIB 
four-form potential $C^{(4)}$ on the basis $\{\gamma_i\}$ 
of harmonic $(1,1)$ forms
\be\label{eq:kKansatzO} 
C^{(4)} = \sum_{i=1}^{k+1} B^i \wedge \gamma_i.
\ee
The tensor fields $B=B^i$ transform in the fundamental 
representation of the global symmetry group $O(k,1)$, 
\[ B^i \to U^i_j B^j, \qquad U \in O(k,1),\]
and are subject to certain self-duality constraints 
formulated in terms of the $O(k,1)$-invariant 
tensor fields  
\[
K^a = x^a_b dB^b + x^a_{k+1}dB^{k+1} \qquad 
H = v_a dB^a + v_{k+1} dB^{k+1}.
\]
$H$ is required to be self-dual and the $K^a$, $a=1,\ldots, k$ 
are required to be anti-self-dual. 
The expectation values of the scalar components $v_a$, $a=1,\ldots,k$ 
and $v_{k+1}$ 
are related to the \kah\ moduli of the F-theory base.
This follows from the fact that the space $H^{1,1}_+(B)$ of 
self-dual $(1,1)$ harmonic forms is spanned by the \kah\ 
class $J_B$. The space of anti-self-dual $(1,1)$ harmonic forms 
is the orthogonal complement of $J_B$ in $H^{1,1}(B)$. 
Let us write the \kah\ class of $B$ as 
\[ 
J_B = t_i \eta^i 
\] 
where $t^i$ are real valued \kah\ moduli. 
Then we have 
\[
H = {1\over 2\mathrm{vol}(B)}\, t_i dB^i.
\]
Note that the volume of the base is parameterized by the expectation value
of a scalar component of a six dimensional hypermultiplet, therefore the 
$t_i$ will be subject to a constraint of the form 
\be\label{eq:kahconstrA}
d^{ij}t_it_j = 2\mathrm{vol}(B) = \mathrm{constant}
\ee
which is reminiscent of the more familiar cubic constraint in five 
dimensional supergravity. By rescaling the fields we may take this constant
to be $1$. Then we can identify $t^i=v^i$, $i=1,\ldots, k+1$, and the
constraint is part of the the orthogonality condition  
\be\label{eq:orthconstrA}
\eta V^T \eta = V^{-1} 
\ee
where $\eta$ is the Minkowski metric tensor of signature 
$(k,1)$.  

For future reference, let us consider the case $k=1$ in more detail. 
In this case, the field $V$ can be chosen of the form
\cite{Romans:1986er}
\be\label{eq:onecase}
V = \left[\begin{array}{cc} \mathrm{cosh}(\phi) & 
\mathrm{sinh}(\phi) \\ \mathrm{sinh}(\phi) & 
 \mathrm{cosh}(\phi) \\ \end{array} \right]
\ee
and the self-dual and anti-self-dual field strengths are given by  
\[ 
\begin{aligned} 
 H & = \mathrm{cosh}(\phi) B^2 + \mathrm{sinh}(\phi) B^1 \cr
K & = \mathrm{cosh}(\phi) B^1 + \mathrm{sinh}(\phi) B^2.\cr
\end{aligned}
\]
This allows us to identify the \kah\ parameters of the 
base $B=\IF_1$ as 
\be\label{eq:fonekah}
t_1= \mathrm{sinh}(\phi)\qquad 
t_2= \mathrm{cosh}(\phi).
\ee
Note that in the case $k=1$ we can give a conventional 
lagrangian formulation of the theory in terms of 
either $B^1+B^2$ or $B^1-B^2$ regarded as an unconstrained 
tensor fields. 

The above considerations are valid for $B=dP_k$, $0\leq k \leq 8$. 
The case $B=\IF_0$ also results in a low energy effective action 
with one tensor multiplet, which has the same tree level formulation
as the case $B=\IF_1$. The main difference between $\IF_0$ and 
$\IF_1$ resides in the relation between the Kaluza-Klein zero modes 
of $C^{(4)}$ and the elementary tensor fields $B^1,B^2$. 
In this case we have 
\be\label{eq:KKansatzB}
C^{(4)} = \left[B^1 \wedge {b-a\over \sqrt{2}} + 
B^2 \wedge {b+a \over \sqrt{2}}
\right].
\ee
The theory can be alternatively formulated in terms of the Kaluza-Klein modes 
$C^1,C^2$ defined with respect to the natural basis $\{a,b\}$ of $(1,1)$ 
forms on $\IF_0$ given by the two rulings,
\be\label{eq:KKansatzC} 
C^{(4)} = C^1 a + C^2 b.
\ee 
Note that 
\be\label{eq:KKansatzD}
C^1 = {B^2-B^1\over \sqrt{2}} \qquad 
C^2 = {B^2+B^1\over \sqrt{2}}. 
\ee
Either $C^1$ or $C^2$ can be regarded as 
unconstrained tensor fields, leading to a conventional lagrangian 
formulation of the theory. 
Moreover, the \kah\ class has the form 
\[
J_B = t_1b + t_2a 
\]
where 
\be\label{eq:fokah}
t_1={1\over \sqrt{2}}e^{\phi} \qquad t_2={1\over \sqrt{2}}e^{-\phi}. 
\ee

The black strings we are interested in are obtained by wrapping D3-branes 
on curves of the form $C=q_i \eta^i$ in $B$, which are charged with 
respect to the elementary tensor fields $B^1,B^2$. 
The charge lattice is 
\[
\Gamma_B \simeq H^{2}(B,\IZ)
\]
equipped with the symmetric bilinear form defined in \eqref{eq:oneonebasisC}.
Charge quantization breaks the global $O(k,1)$ symmetry
group to an integral subgroup $\mathrm{Aut}(\Gamma_B)\subset O(k,1)$. 
In addition, these strings 
carry $n$ units of KK momentum on circle $S^1$ transverse to $B$ and
have angular momentum $J$. The extra charges $(n,J)$
are invariant under U-duality transformations.  

In order to compute the macroscopic energy we have to find
supersymmetric black string solutions of $N=1$ six dimensional 
supergravity with charges $({\underline q}, n, J)$. 
These solutions have been completely classified for 
for the minimal theory (i.e. $k=0$) in
\cite{Gutowski:2003rg} and for gauged supergravity with one 
tensor multiplet (i.e. $k=1$) in \cite{Cariglia:2004kk}.
One can also obtain the results in the ungauged case either by
adapting the results of \cite{Cariglia:2004kk} to the ungauged case,
or, as we will show below, by dualizing solutions of $U(1)^3$ ungauged
supergravity in 5 dimensions.
Analogous results for higher numbers of tensor multiplets 
do not seem to be available in the literature, but an exhaustive 
classification is not really needed for our purposes. 

A very useful observation is that U-duality transformations, which correspond 
to automorphisms of the charge lattice, map supergravity 
solutions to supergravity solutions preserving the entropy. 
Therefore for any $k\geq 2$ we can reduce the problem to $k=1$ 
as long as the charge vector 
\be\label{eq:dpcharges}{\underline q} = q_i \eta^i \ee
can be mapped by a U-duality transformation to a charge vector contained in a 
$(1,1)$ sublattice. For the type of lattices under consideration, this is 
not always the case \cite{Wall}, but we will restiuct ourselves only to such 
charge vectors from now on. 
A similar argument was previously used in a similar context in 
\cite{Maldacena:1999bp}. 
Without loss of generality we can take the $(1,1)$ 
sublattice to be spanned by $(h,e_1)$.

The case $B=\IF_0$ can be easily solved observing that the resulting 
$N=1$ theory expressed for example in terms of the unconstrained 
field $C^2$ is identical to a subsector of the extended $N=4$ supergravity 
obtained by reduction of the IIB theory on $T^4$. The bosonic components 
of the subsector in question are the metric tensor, the six dimensional 
reduction of the RR two-form potential 
$C$ and the dilaton field $\phi$. We will refer to this truncation as the 
D1-D5 subsector since these are precisely the fields which couple with
six dimensional D1-D5 strings. The identification of these two models 
is justified by the isomorphism 
\[
\begin{aligned}  
H^{1,1}(\IF_0) & {\buildrel \simeq \over \longrightarrow}   
H^0(T^4)\oplus H^4(T^4)\cr
(a,b) & \longrightarrow (1,w) \cr
\end{aligned} 
\]
where $w$ is a generator of $H^4(T^4)$ normalized so that 
$\int_{T^4} w =1$. 
This isomorphism is compatible with the bilinear intersection forms. 
Then one can check that the two low energy effective actions are identical 
if we identify $C^2$ with the RR two-form $C$ and the field $e^\phi$
introduced in \eqref{eq:onecase} with the dilaton field. Note that this 
is only a formal identification of the tree level supergravity actions. 
It does not imply that the two physical theories are equivalent, which 
is clearly not the case, but it is a useful technical tool in writing 
down supergravity solutions. In particular note that although the low 
energy fields are formally identified, they have very different 
interpretations in the two theories. For example $e^\phi$ is the dilaton
field in the IIB theory on $T^4$, while it is related to the \kah\ 
parameters of the base in F-theory on $\IF_0$. In the following we will 
think of the D1-D5 subsector of IIB supergravity on $T^4$ just as an auxiliary 
model with no direct physical relevance. 

The identification observed in the last paragraph is useful because 
now one can simply reinterpret the six dimensional solution for a 
D1-D5 string on $T^4$ as a black string solution in the F-theory 
compactification. In particular, the charges of the two solutions 
are related by 
\be\label{eq:chargeidA}
\begin{aligned} 
Q_1 = q_2 \qquad Q_5 = q_1. & \cr
\end{aligned}
\ee
where $q_1,q_2$ are the black string charges with respect to the tensor 
fields $C^1,C^2$ defined in \eqref{eq:KKansatzC}

Since the $\IF_1$ theory is related at tree level to the $\IF_0$ 
by a field redefinition given in equations \eqref{eq:KKansatzB}, 
that we can obtain similarly black string solutions for F-theory 
on $\IF_1$. In this case the charges should be related as follows 
\be\label{eq:fone} 
Q_1 = {q_2+q_1\over \sqrt{2}} \qquad 
Q_5 = {q_2-q_1\over \sqrt{2}}.
\ee
Note that $Q_1,Q_5$ need not be integral since we are only using the 
six dimensional tree level supergravity solution of the D1-D5 system 
as convenient technical tool. At this level, we can simply regard $Q_1,Q_5$ 
as continuous parameters of the solution. 

The case $B=\IP^2$ is somewhat special since it leads to 
minimal $N=1$ supergravity without tensor multiplets.
In fact a black string solution in the $\IP^2$ theory 
can be regarded as a similar solution in the $\IF^1$ theory 
with $q_1=0$. Therefore it will be obtained from the D1-D5 
solution setting 
\be\label{eq:chargeidB}
Q_1=Q_5={q\over \sqrt{2}}
\ee
where ${\underline q} = qh$ is the charge vector of the F-theory 
black string. 

As explained in section two, in our case the black strings wrap a
circle of radius $R$, and are also transverse to a Taub-NUT space.  In
addition to the charges ${\underline q}$ they also carry $n$ units of
KK momentum on this circle and have an angular momentum $J$.
\footnote{ When the transverse space is Taub-NUT, the quantity $J$ is
  not strictly speaking an angular momentum. However, if one replaces
  the transverse Taub-NUT by $\IR^4$ (or if one zooms in near the
  center of a Taub-NUT space of charge one to recover a solution in $\IR^4$)
  this string becomes the six-dimensional lift of a five-dimensional
  BMPV black hole with angular momenta $J_1=J_2=J$. When the transverse
  space is Taub-NUT, the four-dimensional interpretation of the
  quantity $J_1+J_2 = 2 J=m$ is that of D0 charge, or 
KK momentum charge along the  Taub-NUT circle.} The formal 
identification of the corresponding
supergravity solutions to the solution of a D1-D5 system can be
trivially extended to this case.  We will need therefore to find
solutions for a D1-D5 system in an identical six dimensional
background geometry with the same KK momentum $n$ on the circle and
the same angular momentum $J$.

This solution can in fact be obtained by dualizing a five-dimensional
supergavity solution corresponding to M-theory on $T^6\times
TN_r\times R$ with $Q_1, Q_5$ and $n$ M2 branes wrapping three
orthogonal two-cycles in $T^6$ (see for example
\cite{Bena:2005ay,Elvang:2004ds}). Such five-dimensional supergravity
solutions have been classified and studied in much detail in
\cite{Gutowski:2004yv,Bena:2004de,Gauntlett:2004qy}\footnote{See
  \cite{Elvang:2005sa,Gaiotto:2005xt,Bena:2005ni,Chamseddine:1996pi,
    Tseytlin:1996as,Cvetic:1996xz,Breckenridge:1996is,
    Breckenridge:1996sn,
    Sabra:1997yd,Chamseddine:1998yv,Chamseddine:1999qs,Gutowski:2004bj,
    Bena:2004tk,Gauntlett:2004wh,Gutowski:2005id, Bellorin:2006yr} for
  related studies.} and the explicit T-duality transformation can be found for
example in \cite{Bena:2005ay}. 

Let $u$ denote an angular coordinate on $S^1$ with periodicity 
$u \sim u + 2\pi R$ and let $(\psi,x^1,x^2,x^3)$ denote coordinates 
on the Taub-NUT space of charge $r$. 
The angular coordinate $\psi$ has periodicity 
$\psi\sim 4\pi r$, and $(x^1,x^2,x^3)$ 
are cartesian coordinates on the $\IR^3$ base of Taub-NUT. The  
metric is
\be\label{eq:TaubNUT} 
ds_{TN_r}^2 = V d{\vec x}^2 + V^{-1}(d\psi + {\vec A} d{\vec x})^2 
\ee
where 
\[
V = h + {r\over |{\vec x}|}
\] 
and 
\[
\nabla \times {\vec A} = \nabla V.
\] 
The constant $h$ is a modulus that is inversely proportional to the
radius of the circle fiber at infinity. When $h=0$ the metric
(\ref{eq:TaubNUT}) becomes that of $\IR^4$, and the radius in $\IR^4$
is related to $\vec x$ via $r^2_{\IR^4} = 4|{\vec x}| $.

The six dimensional metric and background fields depend on four
harmonic functions \cite{Bena:2005ay,Gauntlett:2004qy}
\be\label{eq:sugrasolA}
\begin{aligned}
& Z_1 = 1 + h c_1+ {Q_1 \over 4|{\vec x}|}& &\qquad& 
& Z_5 = 1 + h c_5 + {Q_5 \over 4|{\vec x}|}& \cr
&Z_p= 1 + h c_p +{n\over 4 |{\vec x}|}& &\qquad& 
& Z_J = h c_J + { J\over 4 |{\vec x}|}& \cr
\end{aligned}
\ee 
on the Taub-NUT space, where $Q_1,Q_5$ are D1 and D5 charges
respectively, $n$ is the KK momentum along the circle, $J$ is the ``5D
angular momentum'' that corresponds to KK momentum along the Taub-NUT
direction, and the parameters $c_1,c_5,c_p$ and $c_J$ are moduli of
the solution.  We work in a convention in which $G_6 = {\displaystyle \pi \over 4} 2
\pi R$, and in which the charges that appear in the supergravity solution are
the same as the quantized D-brane charges.
It is easy to see than when 
$h$ is set to zero, these harmonic functions become the harmonic functions that give the BMPV black hole with angular momenta $J_1=J_2=J$.

In order to write down the metric, let us construct a one-form 
\[
\omega = {Z_J \over 2} (d\psi + {\vec A}d{\vec x}) + {\vec \omega} d\vec x
\]
on the Taub-NUT space, where ${\vec \omega}$ depends only on ${\vec x} 
\in \IR^3$ and is determined by 
\be\label{eq:oneform}
 \nabla \times {\vec \omega} = {1 \over 2} \left( V \nabla Z_J - Z_J \nabla V \right).
\ee
The six dimensional metric is of the form 
\be\label{eq:sugrasolB}
\begin{aligned} 
ds_6^2 = H^{-1} Z_p du^2 - 2H^{-1}du(dt+\omega) + H ds_{TN}^2 
\end{aligned}
\ee 
where 
\[
H \equiv (Z_1Z_5)^{1/2}
.
\]
Note that $\vec \omega$ is determined by condition \eqref{eq:oneform} 
only up to a gradient on $\IR^3$, which can be absorbed by a 
redefinition of the time coordinate $t$. Moreover, the field 
strength $d\omega$ of $\omega$ is anti-self-dual on the Taub-NUT 
space by construction. 
The dilaton $e^{\phi}$ 
is
\be\label{eq:sugrasolC}
\begin{aligned}
e^{\phi}  = \left({Z_1\over Z_5}\right)^{1/2}\cr 
\end{aligned}
\ee

Solutions of the form \eqref{eq:sugrasolB}, \eqref{eq:sugrasolC} can
either be obtained by U-duality from a five-dimensional M-theory on
$T^6$ (or $U(1)^3$) supergravity solution of the BMPV black hole in
Taub-NUT, and also as $u$-independent non-twisting solutions in the
formalism of \cite{Gutowski:2003rg,Cariglia:2004kk}. They describe a
six dimensional black string solutions with a horizon of the form
$S^1\times S^3$. Its macroscopic entropy is
\be\label{eq:macroentDD}
S_{IIB} 
= 2\pi \sqrt{{rnQ_1Q_5} - r^2 J^2}
\ee
independent of the values of the moduli $R,h,c_1,c_5, c_p,c_J$. When
$h=0$ and $r=1$, the Taub-NUT space becomes $\IR^4$, and this black
string reduces to the six-dimensional lift of the BMPV black hole.

One can also understand the macroscopic entropy \eqref{eq:macroentDD}
from a four-dimensional perspective, although as we explained in 
Subsection 4.1, the
values of the moduli at the horizon in the solution of interest make
it intrinsically six-dimensional. If one U-dualizes this solution to
one where the three charges correspond to M2 branes wrapping the three
$T^2$'s of the $T^6$, and then further compactifies the Taub-NUT space
along the fiber, one obtains a four-dimensional black hole that has D2
charges $Q_1,Q_2$ and $n$, KK monopole (D6) charge $r$ and KK momentum
(D0) charge $m = 2J$. The entropy of this black hole is again given by
(\ref{eq:macroentDD}) (see for example \cite{Bertolini:2001ns}.) 

Taking into account the charge identification \eqref{eq:chargeidA}, 
\eqref{eq:chargeidB}, it follows that in the cases $B=\IP^2,\IF_0,\IF_1$, 
the macroscopic entropy of an 
F-theory black string is given by 
\be\label{eq:macroentK}
S_{macro} = \pi \sqrt{2rn Q - r^2 m^2} 
\ee
where \[ Q = d^{ij} q_i q_j.\] 
As explained in the paragraph containing
equation \eqref{eq:dpcharges}, the case $B=dP_k$, $2\leq k \leq 8$ 
can be reduced to the case $B=\IF_1$ if the charge 
vector \eqref{eq:dpcharges} can be mapped by an automorphism of $\Gamma_B$
to a $(1,1)$ sub-lattice. Therefore formula 
\eqref{eq:macroentK} will hold in those cases as well. 

Finally, note that the \kah\ moduli of the F-theory base are fixed 
by an attractor mechanism. For the $\IF_0$ model, using equations 
\eqref{eq:fokah}, \eqref{eq:chargeidA}, we find  
\be\label{eq:fomodel} 
t_1 = {1\over \sqrt{2}}{q_2\over q_1} \qquad 
t_2 = {1\over \sqrt{2}}{q_1\over q_2}
\ee
at the attractor point. As expected, these values are independent of 
the moduli of the solution. For the $\IF_1$ model, equations 
\eqref{eq:fonekah} and \eqref{eq:chargeidB} yield 
\be\label{eq:fonemodel}
t_1 = {1\over 2} \left({q_2 \over q_1} -{q_1 \over q_2}\right)\qquad 
t_2 = {1\over 2} \left({q_2 \over q_1} + {q_1 \over q_2}\right).
\ee
Note that the solution is physically sensible only if $t^1,t^2$ are 
positive. For the $\IF_0$ model, this will hold if $q_1,q_2>0$ while 
for the $\IF_1$ model we need  
$q_1,q_2>0$, and $q_2>q_1$. 
These are precisely the ampleness conditions for the divisor 
$C=q_i\eta^i$ on $\IF_0$ and $\IF_1$ respectively.  
For more general models, the values of the \kah\ parameters can be 
obtained by U-duality transformations. 

It is also worth noting that the attractor mechanism also fixes the 
radius of the circle parameterized by $u$ to 
\[
n \sqrt{2\over rQ}.
\]
If we take $n$ much larger than $Q$, the circle is very large at the 
attractor point, hence the geometry is six dimensional. This is consistent 
with the behavior of the \kah\ parameters of the four dimensional 
attractor solutions found in the previous subsection.

\subsection{Comparison with Microscopic Entropy}

Our next goal is to understand the relation between the microscopic 
entropy formula 
\eqref{eq:microent} and the macroscopic formula \eqref{eq:macroentK}. 
Summarizing conditions $(i)$ and $(ii)$ below \eqref{eq:kahmodB}, 
recall that the microscopic entropy formula is reliable if 
\be\label{eq:limit}
n >> q_i >> 0 ,\qquad q_i >> rc_i,
\ee
assuming that $\eta^i$ are generators of the Mori cone of $B$. 
In this limit we have 
\[ Q >> c\] 
since  
\[
Q-c = (C\cdot (C+K_B))_B = 2g(C)-2 
\]
is the arithmetic genus of $C$, which is very large and positive 
for very large $q_i$. 

Let us examine the behavior of the microscopic entropy \eqref{eq:microent} 
in this limit. 
The leading term in the expression of the triple intersection 
\eqref{eq:tripleB} is  
\be\label{eq:tripleC} 
D \sim {rQ\over 2}. 
\ee 
This yields 
\be\label{eq:microentB}
S_{micro} \sim \pi \sqrt{2rQ {\widehat m}}
\ee 
where ${\widehat m}$ is given by \eqref{eq:effdzero}. 
The leading term of ${\widehat m}$ at large $m$ 
is given by 
\be\label{eq:leadingdzero}
{\widehat m} \sim n + {1\over 12} D^{\alpha\alpha} m^2. 
\ee
In order to compute $D^{\alpha\alpha}$, first note that 
\[ 
\begin{aligned}
\mathrm{det}\left[
\begin{array}{cc} 
D_{\alpha \alpha} & D_{\alpha i} \cr
D_{i\alpha } & D_{ij} \cr
\end{array} 
\right] & = \mathrm{det}\left({r\over 6}d_{ij}\right)
\left({c\over 6} - {Q\over 6r}\right)
 \sim -{Q\over 6r}  \mathrm{det}\left({r\over 6}d_{ij}\right).\cr
\end{aligned}
\]
Then we have 
\[ 
D^{\alpha\alpha}\sim -{6r\over Q} 
\]
and \eqref{eq:leadingdzero} becomes 
\[ 
{\widehat m} \sim n - {r\over 2Q}m^2. 
\]
Therefore the leading behavior of the microscopic energy 
\eqref{eq:microentB} is 
\be\label{eq:microentC} 
S_{micro} \sim \pi \sqrt{2rnQ - r^2m^2}
\ee
which is identical to the leading behavior of the macroscopic 
formula \eqref{eq:macroentK}.

\subsection{Subleading Corrections} 

We conclude this section with a brief discussion of subleading 
corrections. 
So far we have taken into account only
the leading terms in the expression of the left moving central charge 
\eqref{eq:leftcharge} in the limit \eqref{eq:limit}. 
There are two types of subleading corrections. One could take into account 
subleading terms in the expression of the triple intersection 
\eqref{eq:tripleB} and the correction terms of the form 
\[ 
{1\over 6}\int_X (r\sigma + \pi^*C) \wedge c_2(X)
\]
to the central charge. Here we will concentrate only on the first 
type of subleading terms, which have the same scaling behavior 
as the leading term \eqref{eq:tripleC} with respect to the charges 
$q_i,r$. Corrections of the second type are linear in the charges, 
hence they have a lower scaling behavior. 

The microscopic formula becomes 
\be\label{eq:microentD} 
S_{micro} \sim \pi \sqrt{2rn\left(Q-cr+{dr^2\over 3}\right)-r^2m^2}.
\ee
The question is if the subleading terms present in 
\eqref{eq:microentD} can be understood from a supergravity 
analysis. 

Let us first try to understand the origin of such corrections in F-theory. 
So far we have been working with tree level $N=1$ supergravity, 
which can be regarded as a truncation of the $N=4$ theory. 
However the low energy description of F-theory has extra 
couplings which are not consistent with a truncation of the 
$N=4$ theory. The couplings in question are six-dimensional 
Green-Schwartz terms required by anomaly cancellation
\cite{Sagnotti:1992vb,Morrison:1996na,Morrison:1996pp,Ferrara:1996wv,
Sadov:1996zm,Ferrara:1997gh}. In this paper we consider only F-theory 
compactifications on smooth elliptic fibrations, therefore we do not 
have six dimensional vector multiplets. The theory will have only 
gravitational anomalies, which determine the higher curvature 
corrections to the tree level supergravity action. 

According to \cite{Sagnotti:1992vb,Ferrara:1996wv,Sadov:1996zm,Ferrara:1997gh}
the higher curvature terms are encoded in a shift of the elementary 
field strengths $H^i =dB^i$ by a gravitational Chern-Simons term. 
More precisely, one has to define 
\be\label{eq:shiftA} 
H^i = dB^i + a_i \omega 
\ee 
where $\omega$ is the gravitational Chern-Simons term for the six dimensional 
spin connection. According to  
\cite{Sadov:1996zm}, the coefficients $a_i$ are given by 
$a_i = {c_i\over 2}$, 
where the $c_i$ were defined in \eqref{eq:notation}.
The effect of this shift on the supersymmetry variations 
and equations of motion has been worked out in 
\cite{Sagnotti:1992vb,Ferrara:1996wv,Ferrara:1997gh}. 
In principle one should solve the new equations of motion and 
BPS conditions in order to understand the effect of higher 
curvature corrections on the black string entropy. This would be 
quite an involved analysis which we will leave for future work. 

However, let us observe that if we ignore the back-reaction of the 
noncritical string on the six dimensional space-time geometry, 
the shift \eqref{eq:shiftA} results in a shift of the form 
\be\label{eq:shiftB} 
q^i \to q^i -{rc_i\over 2} 
\ee
on the charges. This follows by a direct evaluation of the Chern-Simons 
term in a Taub-NUT background. Such a shift is reminiscent of a similar 
modification of black hole charges in the four dimensional attractor 
mechanism \cite{Shmakova:1996nz}. In fact it can be easily checked that this 
is indeed the shift predicted in \cite{Shmakova:1996nz} for the attractor 
solutions discussed in section 4.1. One can think about 
the correction term in \eqref{eq:shiftA} as giving rise to a difference between
the charge measured at infinity, $q^i$, and the actual charge of 
the black hole, $q^i - {\displaystyle r c_i \over 2}$.
Accepting this shift on a conjectural 
basis for the moment, note that it would result in a modified macroscopic 
entropy formula of the form 
\be\label{eq:macroentL} 
S_{macro} = \pi \sqrt{2rn\left(Q-cr+{dr^2\over 4}\right)-r^2m^2}. 
\ee
Quite remarkably, this formula exhibits the same subleading 
correction as the microscopic result \eqref{eq:microentC}, 
but the next order corrections, namely the terms proportional to 
$nr^3d$, are different. These terms are very small in the limit 
\eqref{eq:limit}, but they would become important in a regime 
in which $q_i$ and $rc_i$ are of the same order of magnitude. 
This is precisely the regime in which we also expect the effects
of the singular divisors on the microscopic entropy formula 
to be become important, as explained below \eqref{eq:kahmodB}.
It would be very interesting to confirm the macroscopic entropy 
formula \eqref{eq:macroentL} by a direct supergravity computation. 
If the result conjectured here is indeed valid, it would also be 
very interesting to understand the microscopic computation in the 
regime $q_i\sim rc_i$ and compare the two expressions.

\bibliography{elliptic.bib}
\bibliographystyle{utcaps}

\end{document}